\documentclass[twocolumn,prb,showpacs]{revtex4-1}
\usepackage{times,xspace}
\usepackage{graphicx,graphics,color,epsfig}
\usepackage{bm}
\usepackage{amsmath}
\usepackage{amssymb}
\usepackage{epstopdf}
\usepackage{dcolumn}

\newcommand{\blue}{\textcolor{black}}

\begin{document}

\title{Testing the sign-changing superconducting gap in iron-based superconductors with quasiparticle interference and neutron scattering}

\author{Tanmoy Das$^1$, and A. V. Balatsky$^{1,2}$}
\affiliation{$^1$Theoretical Division, Los Alamos National Laboratory, Los Alamos, NM, 87545, USA.\\
$^2$Center for Integrated Nanotechnologies, Los Alamos National Laboratory, Los Alamos, NM, 87545, USA.}
\date{\today}
\begin{abstract}
We present a phenomenological calculation of the quasiparticle-interference (QPI) pattern and inelastic Neutron scattering (INS) spectra in iron-pnictide and layered iron-selenide compounds by using materials specific band-structure and superconducting (SC) gap properties. As both the QPI and the INS spectra arise due to scattering of the Bogolyubov quasiaprticles, they exhibit an one-to-one correspondence of the scattering vectors and the energy scales. We show that these two spectroscopies complement each other in such a way that a comparative study allows one to extract the quantitative and unambiguous information about the underlying pairing structure and the phase of the SC gap. Due to the nodeless and isotropic nature of the SC gaps, both the QPI and INS maps are concentrated at only two energies in pnictide (two SC gaps) and one energy in iron-selenide, while the associated scattering vectors ${\bf q}$ for scattering of sign-changing and same-sign of the SC gaps change between these spectroscopies. The results presented, particularly  for newly iron-selenide compounds, can be used  to test the nodeless $d-$wave pairing in this class of high temperature superconductors.
\end{abstract}
\maketitle \narrowtext

\section{Introduction}
The most crucial information to unravel the mechanism of pairing is the structure of
the superconducting (SC) gap function, a measure of the amplitude and the phase of electron pairs. Although the SC-gap function of conventional phonon-mediated (attractive pairing interaction) superconductors has the same sign all over the $k-$space ($s-$wave symmetry), that of spin-fluctuation mediated (repulsive pairing interaction) superconductors, within the weak to intermediate coupling scenario,\cite{strongcoupling} is expected to exhibit a sign reversal between the Fermi momenta connected by the characteristic `hot-spot' wave vector ${\bf Q}$ of spin-fluctuations.\cite{scalapino,Pines} As a consequence of the sign reversal, nodal planes in which the SC gap vanishes should exist in $k$ space. If such nodal planes intersect the Fermi surface (FS), as in $d-$wave cuprate, nodal quasiparticle states emerge in the low-energy spectra. However, if the FS consists of small pockets and do not touch the nodal line, as $s^{\pm}$ pairing in iron-pnictide and single layer iron-chalogenides\cite{Mazin} or $d$-wave pairing in double layered iron-selenide (Fe$_2$Se$_2$),\cite{DHLeeFeSe,MaierFeSe,DasFeSe,Dasvacancy} develop despite a sign reversal of the pairing. Therefore, to precisely establish an unconventional pairing symmetry and separate it from a conventional $s-$wave, the relative sign of the SC gap on the FS must be established.

The magnetic resonance mode develops in the SC state and was shown theoretically to arise from the sign-flip of the SC gap at the `hot-spot'.\cite{chubukov,norman,eremin,Dastworesonances,DasFeSe} However, there are other theories
which also reproduce the resonance mode without the requirement of sign change of the SC gap.\cite{so5,PLee_slaveboson,wilsonPRB} Alternative and important new spectroscopy comes from the quasiparticle interference (QPI) pattern, measured by scanning tunneling microscopy (STM), which in principle visualizes all possible elastic scattering vectors. It therefore can distinguish the sign-changing `hot-spot' vectors by studying its evolution with varying magnetic field.\cite{Hanaguri_cuprate,Hanaguri_pnictide,knolleprl,knolleprb} Magnetic field breaks the time-reversal symmetry of the SC quasiparticles and thus illuminate those scattering vectors which scatter quasiparticles of same pairing phase, thanks to the remarkable properties of the Bogolyubov coherence factors.\cite{Schrieffer_book} These observations can be questioned based on  at least two arguments: (1)  magnetic field induced vortex states drastically redistribute the spectral weight across the `bright-spots' on the constant energy surface, making this procedure complicated; (2)  a magnetic impurity always carries a scalar potential which allows simultaneous scattering of quasiparticles of opposite pairing phase.\cite{knolleprl,knolleprb}  We show in this paper that the INS and QPI maps complement each other and allow determination of the relative phase. Thus, we propose,  a comparative study between the Inelastic neutron Scattering (INS) spectra and the QPI pattern, taken {\em together}, can be a viable  tool to quantitatively and unambiguously determine the relative sign of the SC gap.

Our approach is based on a very simple observation: Both the INS and the QPI patterns arise from the similar scattering of the Bogoliubov quasiparticles, forming  Cooper pairs (inelastic and elastic scatterings, respectively) and thus their observed dispersions, $\Omega({\bf q})$s, must have an one-to-one relationship in these two spectroscopies. In INS spectra, a scattering ${\bf q}$ vector will be observed if it connects two quasiparticle states at ${\bf k}_i$ and ${\bf k}_f$ at which ${\rm sgn}\bigl[\Delta_{{\bf k}_i}\bigr]\ne{\rm sgn}\bigl[\Delta_{{\bf k}_f}\bigr]$ and thus its energy scale will be determined by $|\Delta_{{\bf k}_i}|+|\Delta_{{\bf k}_f}|$. The same vector will only appear in the QPI spectra if $|\Delta_{{\bf k}_i}|=|\Delta_{{\bf k}_f}|$, because QPI  probes elastic scattering. The comparison between the INS and the QPI maps can therefore emphasize the regions of momentum space with opposite sign of the gap function,  connected by scattering momenta. This emphasis can be  further facilitated by studying the magnetic field dependence of the QPI maps. Such a comparative study has proven to give valuable information about the pairing symmetry in cuprates.\cite{electron_pocket}

{\it Iron-pnictide:-} In the case of iron-pnictide superconductors, however, the presence of multiple bands at the Fermi level ($E_F$) and multiple SC gap amplitudes at each bands make the aforementioned analysis much more complicated and exotic. In these compounds, the FS consists of disconnected two concentric hole pockets (namely, $\alpha$ and $\beta$ pockets) and one electron pocket ($\gamma$ pocket) centered at $\Gamma$ and M points, respectively. Prominent nesting between the hole pockets and the electron pocket has been shown theoretically to lead to a $s^{\pm}-$pairing in this family of superconductors.\cite{Mazin} Within this pairing, all the FS pockets possess nodeless and isotropic SC gap, but the sign of the gaps is reversed between the hole pockets and the electron pocket. To complicate the story, the two hole pockets acquire very different magnitude of the SC gaps.\cite{HDing_twogaps,Johnston,Dastworesonances} Therefore, macroscopic geometrical separation of the two phases, as in the cuprate corner-junction experiment, can not be performed in these systems to test the pairing symmetry.\cite{Hoffman}

In our earlier study, we have theoretically shown that because of the two values of the SC gap in two hole-pockets, the INS spectra is split into two energy scales.\cite{Dastworesonances} Furthermore, as the two hole pockets have very different area as a function of doping, the corresponding ${\bf q}$ vectors are also different at these two resonances. Here, we show that these INS scattering vectors also appear in two different energy scales of the QPI patterns. Additional scattering vectors between the states of same sign of SC gaps, which were prohibited in INS, appear in QPI. These results are consistent with STM data on electron doped Ba(Fe$_{1-x}$Co$_x$)$_2$As$_2$ as a function of doping.\cite{NCYeh} Furthermore, the evolution of the QPI pattern as a function of magnetic field is also consistent with STM studies in single layer Fe(Se,Te) superconductors.\cite{Hanaguri_pnictide}

{\it Fe$_2$Se$_2$ superconductors:-} The recent discovery of high-$T_c$ superconductivity in double layered Fe$_2$Se$_2$ based superconductors makes the above story even more interesting and exotic.\cite{FeSe,FeSe1,FeSe2} The FS in this family of superconductors only hosts electron pockets at M point (no hole-pocket is present here as in iron-pnictide discussed above). Therefore, the leading nesting vector occurs between the two electron pockets which lead to a $d$-wave pairing ($d_{x^2-y^2}$ in 1 Fe unit cell and $d_{xy}$ in 2 Fe unit cell notation).\cite{Dastworesonances} Unlike the $d$-wave pairing in cuprates, here it gives rise to nodeless and isotropic SC gap on the FS, in consistent with experiments\cite{Zhang,HDing,Mou,Cv}, but  gap function changes sign between the two electron pockets. We and others \cite{DHLeeFeSe,MaierFeSe,DasFeSe} have shown earlier that such a $d$-wave pairing leads to a magnetic resonance mode at ${\bf Q}$ which is observed later by INS measurements.\cite{Inosov_INS_FeSe}

Here we show  that this INS scattering vector ${\bf Q}$  will also show up in the QPI vector at the bias energy equals to the SC gap magnitude. Furthermore, we phenomenologically demonstrate that with the application of the magnetic field, other QPI scattering vectors at which the SC gap do not change sign can be illuminated. The relative evolution of the QPI maps as a function of magnetic field will provide a valuable test for the relative phase of the SC state on the FS in this family of superconductors.

The paper is organized as follows. In Sec. II, we provide the formalism for both QPI and INS spectra. We extract out very simple equations for these two spectroscopies which can be used with experimental inputs to reconcile them. In Sec. III, we present the results for iron-pnictide, whereas the same results but for layered iron-selenide are given in Sec. IV. The magnetic field dependence of the QPI map is computed for layered iron-selenide system in Sec.~IVA. Finally, we conclude in Sec. V.

\section{Formalism}
A direct correlation between the INS and the QPI spectra can only be made within the bare level in which the same green's functions are involved. This correspondence therefore would at least qualitatively be correct for dressed Greens functions. In fact, one can again evaluate the Green's functions for the filled state from ARPES spectral weight $A$, as $G({\bf k},i\omega_n)=\pi\int A({\bf k},i\omega^{\prime})/(i\omega_n-\omega^{\prime})$, which is exact for a single band case, but averaged over orbital indices in a multiband system. In a multiband superconductor, the Green's functions for the normal and anomalous part can be written in the eigenbasis as
\begin{eqnarray}
G_{\nu}({\bf k},i\omega_n)&=&\frac{\alpha^2_{\nu}({\bf k})}{i\omega_n-E_{\nu}({\bf k})}+\frac{\beta^2_{\nu}({\bf k})}{i\omega_n+E_{\nu}({\bf k})},\\
F_{\nu}({\bf k},i\omega_n)&=&\alpha_{\nu}({\bf k})\beta_{\nu}({\bf k})\nonumber\\
&&~\times\left(\frac{1}{i\omega_n-E_{\nu}({\bf k})}-\frac{1}{i\omega_n+E_{\nu}({\bf k})}\right),
\end{eqnarray}
respectively. $n$ is the Matsubara frequency. Here $E_{\nu}({\bf k})=\pm\sqrt{\xi_{\nu}^2({\bf k})+\Delta_{\nu}^2({\bf k})}$ is the $\nu^{th}$ Bogolyubov quasiparticle band. $\xi_{\nu}^2({\bf k})$ is the non-interaction band modeled by tight-binding parametrization to the material specific LDA bands, as discussed later. $\Delta_{\nu}({\bf k})=\Delta^0_{\nu}g({\bf k})$ is the momentum dependent SC gap with $\Delta^0_{\nu}$ is the band specific SC gap magnitude. $g({\bf k})$ is the structure factor for the pairing symmetry which is taken to be same for all bands in a given system. $\alpha_{\nu}({\bf k}) (\beta_{\nu}({\bf k}))=\sqrt{\frac{1}{2}\left(1\mp\frac{\xi_{\nu}({\bf k})}{E_{\nu}({\bf k})}\right)}$ are the Bogolyubov coherence factors for the quasiparticle states $\pm E_{\nu}({\bf k})$, respectively. The above Green's functions can be projected to the orbital basis as $G_{sp}({\bf k},i\omega_n)=\bigl<\phi_{\nu}^s({\bf k})|G_{\nu}({\bf k},i\omega_n)|\phi_{\nu}^p({\bf k})\bigr>$ and same for $F$. $\phi_{\nu}^s({\bf k})$ represent the eigenvectors of $\nu^{th}$ band onto $s^{th}$ orbital.

{\it QPI profile:-}STM measures local density of state (LDOS) of quasiparticles which arrive to the tip after going through multiple intrinsic elastic scattering (due to magnetic and non-magnetic scatterer\cite{SashaQPI,knolleprl,knolleprb}) in the system. Such scattering vectors can be visualized by Fourier transforming the LDOS into the ${\bf q}$ space to obtain\cite{BobQPI,SashaQPI,Dastwogapcuprate,knolleprl,knolleprb,ZhangQPI}
\begin{eqnarray}\label{eq:QPI}
&&B_{rstu}({\bf q},i\Omega_n)={\rm Im}\sum_{{\bf k},\nu,\nu^{\prime}}M^{\nu\nu^{\prime}}_{rstu}({\bf k},{\bf q})\nonumber\\
&&~~~\times |V_{\bf q}|\bigl[G^{\nu}({\bf k},i\Omega_n)G^{\nu^{\prime}}({\bf k}+{\bf q},i\Omega_n)\nonumber\\
&&~~~~~~-F^{\nu}({\bf k},i\Omega_n)F^{\nu^{\prime}}({-{\bf k}-{\bf q}},-i\Omega_n)\bigr].
\end{eqnarray}
Here $M$ is the orbital to band matrix element made of the eigenvectors as $M^{\nu\nu^{\prime}}_{rstu}({\bf k},{\bf q})=\phi_{\nu}^{r*}({\bf k} +{\bf q})\phi_{\nu^{\prime}}^{s}({\bf k})\phi_{\nu^{\prime}}^{t*}({\bf k})\phi_{\nu}^{u}({\bf k}+{\bf q})$. The matrix-element can be important for shaping the spectral weight distributions for multiband case. Nevertheless, the relative intensity of the scattering vectors is determined by the nesting conditions and the SC gap amplitude for elastic scattering. In addition, a scattering matrix-element, $C({\bf k},{\bf q})$, appears due to the coherence factors, $\alpha_{\nu}({\bf k})$ and $\beta_{\nu}({\bf k})$, of the Bogolyubov quasiparticles which is sensitive to the momentum-dependent phase of the SC order parameter and the symmetry of the scattering  potential, $V_{\bf q}$.\cite{Schrieffer_book,Hanaguri_cuprate} The green's function is expressed in terms of the spectral weight as $G^{\nu}({\bf k},i\Omega_n)=-\int_{-\infty}^{\infty}\frac{d\omega^{\prime}}{2\pi}A^{\nu}({\bf k},\omega^{\prime})/(i\Omega_n-\omega^{\prime})$. If no many-body broadening is included then the spectral weight can be presented by delta function. Taking these facts into account, we simplify Eq.~\ref{eq:QPI} to calculate the total QPI map as,\cite{Hanaguri_cuprate}
\begin{eqnarray}\label{eq:QPI2}
B({\bf q},\Omega_{\rm QPI})&\approx&|V_{\bf q}|\sum_{{\bf k},\nu,\nu^{\prime}} C_{\nu\nu^{\prime}}({\bf k},{\bf q}) \nonumber\\
&&\times\delta\bigl(\Omega_{\rm QPI}-E_{\nu}({\bf k})\bigr)\delta\bigl(\Omega_{\rm QPI}-E_{\nu}({\bf k}+{\bf q})\bigr).\nonumber\\
\end{eqnarray}
[Here we assumed $M=1$ for simplicity\cite{ZhangQPI} and took analytical continuation $i\Omega_n\rightarrow\Omega_{\rm QPI}+i\eta$ ($\eta$ is a small broadening).] The explicit form of the scattering matrix element as given in Ref.~\onlinecite{Hanaguri_cuprate,Hanaguri_pnictide} is $C_{\nu\nu^{\prime}}({\bf k},{\bf q})=
\bigl({\rm sgn}[\Delta_{\nu}({\bf k})]{\rm sgn}[\Delta_{\nu^{\prime}}({\bf k}+{\bf q})]\alpha_{\nu}({\bf k})\alpha_{\nu^{\prime}}({\bf k}+{\bf q})\mp \beta_{\nu}({\bf k})\beta_{\nu^{\prime}}({\bf k}+{\bf q})\bigr)^2$, where the negative and positive signs represent scattering through a scalar (even under time-reversal) and a magnetic (odd under time-reversal) potential, respectively. Furthermore, scattering from disorder that converts electrons into holes as they are scattered, gives rise to the coherence factor  $C_{\nu\nu^{\prime}}({\bf k},{\bf q})=
\bigl({\rm sgn}[\Delta_{\nu}({\bf k})]\alpha_{\nu}({\bf k})\beta_{\nu^{\prime}}({\bf k}+{\bf q}) + {\rm sgn}[\Delta_{\nu^{\prime}}({\bf k}+{\bf q})]\beta_{\nu}({\bf k})\alpha_{\nu^{\prime}}({\bf k}+{\bf q})\big)\bigl({\rm sgn}[\Delta_{\nu}({\bf k})]{\rm sgn}[\Delta_{\nu^{\prime}}({\bf k}+{\bf q})]\alpha_{\nu}({\bf k})\alpha_{\nu^{\prime}}({\bf k}+{\bf q})+ \beta_{\nu}({\bf k})\beta_{\nu^{\prime}}({\bf k}+{\bf q})\bigr)$. At $E_F$ the two delta function in Eq.~\ref{eq:QPI2} gives $\Omega^{\nu\nu^{\prime}}_{\rm QPI}({\bf q})=|\Delta_{\nu}({{\bf k}_F})\bigr|=\bigl|\Delta_{\nu^{\prime}}({\bf k}_F+{\bf q})\bigr|$. Imposing this constraint, we obtain $\alpha_{\nu}({\bf k_F})=\alpha_{\nu^{\prime}}({\bf k_F}+{\bf q})$, and $\beta_{\nu}({\bf k_F})=\beta_{\nu^{\prime}}({\bf k_F}+{\bf q})$ at $E_F$ (by substituting $\xi_{k}=0$ in the equations for $\alpha,\beta$ given above). This leads to a systematic selection rule for  ${\bf q}$ vectors which depends on the nature of the scatterer: In the case of weak scalar potential scattering, $C\sim 0$ for those ${\bf q}$ at which ${\rm sgn}\bigl[\Delta_{\nu}({\bf k}_F)\bigr]={\rm sgn}\bigl[\Delta_{\nu^{\prime}}({\bf k}_F+{\bf q})\bigr]$. By contrast, for scattering off magnetic impurities or gap inhomogeneities, $C\sim 0$ for those ${\bf q}$ at which ${\rm sgn}\bigl[\Delta_{\nu}({\bf k}_F)\bigr]\ne{\rm sgn}\bigl[\Delta_{\nu^{\prime}}({\bf k}_F+{\bf q})\bigr]$. Therefore, focusing on the low-energy region where $\Omega_{\rm QPI}\le\Delta$, we can write the conditions to obtain a non-vanishing QPI vector as
\newline\\
\begin{centering}
\begin{tabular}{|p{9.cm}|}
\hline
\begin{eqnarray}
&&\Omega^{\nu\nu^{\prime}}_{\rm QPI}({\bf q})=|\Delta_{\nu}({{\bf k}_F})\bigr|=\bigl|\Delta_{\nu^{\prime}}({\bf k}_F+{\bf q})\bigr|,
\label{eq:QPI3}\\
&&{\rm sgn}\bigl[\Delta_{\nu}({\bf k}_F)\bigr]\ne{\rm sgn}\bigl[\Delta_{\nu^{\prime}}({\bf k}_F+{\bf q})\bigr]~{\rm for~scalar~imp.,}
\label{eq:QPI4}\\
&&{\rm sgn}\bigl[\Delta_{\nu}({\bf k}_F)\bigr]={\rm sgn}\bigl[\Delta_{\nu^{\prime}}({\bf k}_F+{\bf q})\bigr]~{\rm for~mag.~imp.}
\label{eq:QPI5}
\end{eqnarray}
\\
\hline
\end{tabular}
\end{centering}
\newline\\

Note that as magnetic impurity is always associated with a scalar potential, it, in principle, involves QPI scatterings which satisfies Eq.~\ref{eq:QPI4} as well but its intensity will depend on the relative strength of the potential. We take $V({\bf q})$ as a constant potential.

{\it INS spectra:-}The calculation of the INS spectra follows similarly with the exception that the latter is an inelastic scattering of the quasiparticle spectra. INS probes the imaginary part of the susceptibility which can be written in the orbital basis as\cite{maier,Dastworesonances}
\begin{eqnarray}\label{eq:chi0}
&&\chi_{0rstu}({\bf q},i\Omega_m)=-\frac{1}{2}\sum_{k,n,\nu,\nu^{\prime}}M^{\nu\nu^{\prime}}_{rstu}({\bf k},{\bf q})\nonumber\\
&&~~~\times\bigl[G^{\nu}({\bf k},i\omega_n)G^{\nu^{\prime}}({\bf k}+{\bf q},i\omega_n+i\Omega_m)\nonumber\\
&&~~~~~+F^{\nu}({\bf k},i\omega_n)F^{\nu^{\prime}}({-{\bf k}-{\bf q}},-i\omega_n-\Omega_m)\bigr].
\end{eqnarray}
We have shown earlier in Ref.~\onlinecite{Dastworesonances} that in many cases the difference between realistic matrix element and  matrix element assumed to be a smooth function of energy and momentum and thus is not important in calculating the magnetic scattering structure. The INS dispersion is mostly governed by the locus of the discontinuous jumps in $\chi_0$ and due to Kramers-Kroning relationship, corresponding $\chi_0^{\prime\prime}$ attains a peak at the same location [within random-phase approximation (RPA), the peak position shifts slightly to a lower energy]. Similar to Eq.~\ref{eq:QPI2} above, we absorb the INS scattering matrix-element term in $C_{\nu\nu^{\prime}}({\bf k},{\bf q})$  and performing the Matsubara summation in Eq.~\ref{eq:chi0}, we obtain the total BCS $\chi_0$ as
\begin{eqnarray}
\chi_{0}({\bf q},\Omega_{\rm INS})&\approx&\sum_{{\bf k},\nu,\nu^{\prime}} C_{\nu\nu^{\prime}}({\bf k},{\bf q}) \nonumber\\
&&\times \delta\bigl(\Omega_{\rm INS}-E_{\nu}({\bf k}) -E_{\nu}({\bf k}+{\bf q})\bigr).
\label{eq:chi2}
\end{eqnarray}
Here, the explicit form of $C$ is $C_{\nu\nu^{\prime}}({\bf k},{\bf q})=\beta_{\nu}({\bf k})\alpha_{\nu^{\prime}}({\bf k}+{\bf q})\bigl[\alpha_{\nu}({\bf k})\beta_{\nu^{\prime}}({\bf k}+{\bf q})-\beta_{\nu}({\bf k})\alpha_{\nu^{\prime}}({\bf k}+{\bf q})\bigr]=\bigl(1-{\rm sgn}[\Delta_{\nu}({\bf k}_F)]{\rm sgn}[\Delta_{\nu^{\prime}}({\bf k}_F+{\bf q})]\bigr)$ at $E_F$.

This implies that the magnetic structure in BCS $\chi_0$ below $\Omega_{\rm INS}\le2\Delta$ is entirely governed by two conditions:\cite{chubukov,norman,eremin,Dastworesonances,DasFeSe}
\newline\\
\begin{centering}
\begin{tabular}{|p{8.5cm}|}
\hline
\begin{eqnarray}
&&\Omega^{\nu\nu^{\prime}}_{\rm INS}({\bf q})=\bigl|\Delta_{\nu}({{\bf k}_F})\bigr|+\bigl|\Delta_{\nu^{\prime}}({\bf k}_F+{\bf q})\bigr|,
\label{eq:chi3}\\
&&{\rm sgn}\bigl[\Delta_{\nu}({\bf k}_F)\bigr]\ne{\rm sgn}\bigl[\Delta_{\nu^{\prime}}({\bf k}_F+{\bf q})\bigr].
\label{eq:chi4}
\end{eqnarray}
\\
\hline
\end{tabular}
\end{centering}
\newline\\

We will use QPI Eqs.~\ref{eq:QPI3}-\ref{eq:QPI5} and INS Eqs.~\ref{eq:chi3} \& \ref{eq:chi4} to perform a comparative analysis of the two spectroscopic data in iron-pnitide and iron-selenide superconductors.

\section{Results on Iron-Pnictide}

\begin{figure}
\rotatebox{0}{\scalebox{0.6}{\includegraphics{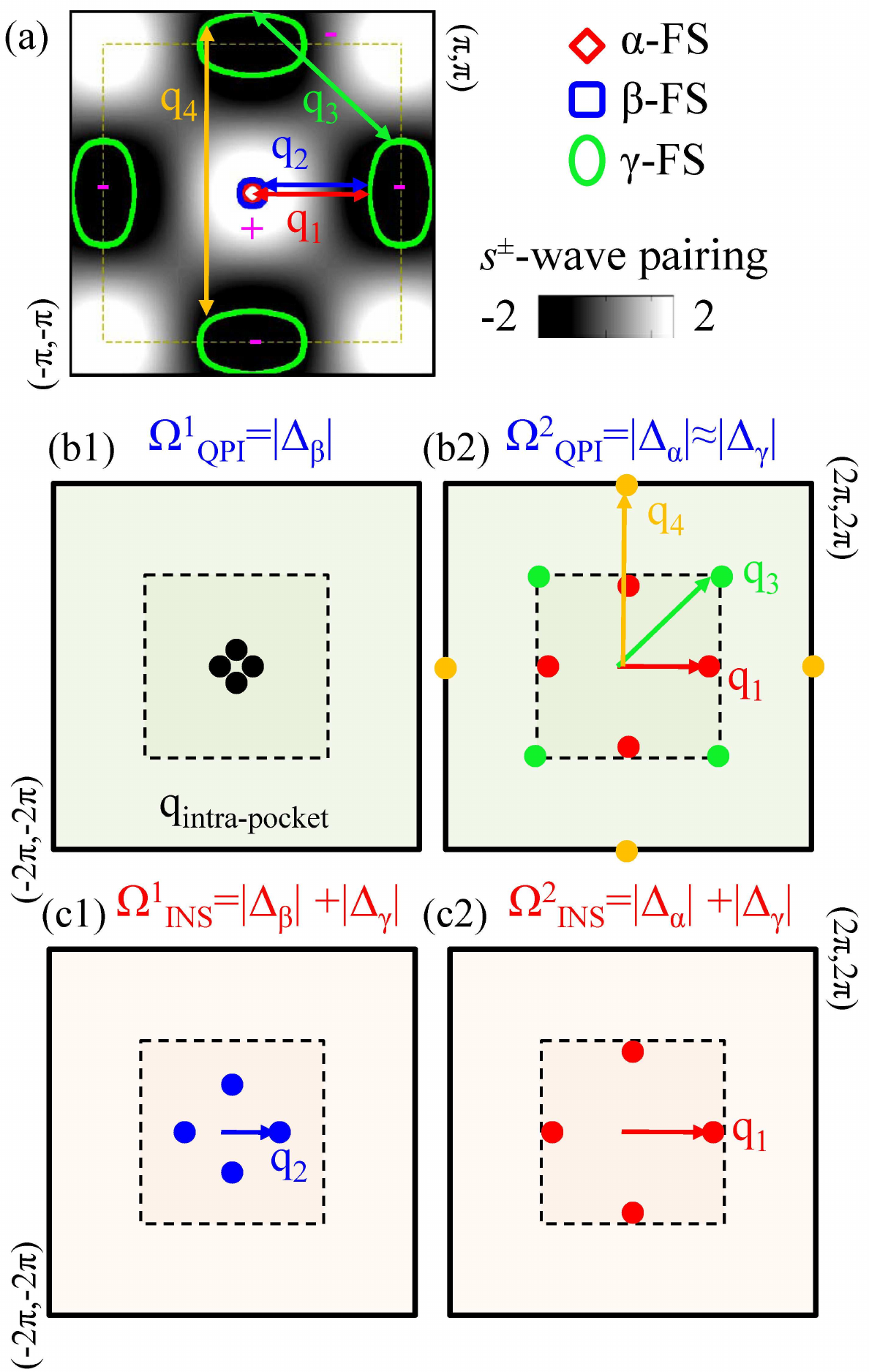}}}
\caption{(a) Computed FS for iron-pnictide within five band tight-binding model at a representative electron doping of $x=0.1$. The BZ is chosen for 1 Fe unit cell notation. The black to white background depicts the $s^{\pm}$-pairing symmetry which takes the form of $2\cos{(k_xa)}\cos{(k_ya)}$ in the 1 Fe unit cell. The arrows give different interband scattering channels which constitute QPI and INS maps. (b1) According to Eq.~\ref{eq:QPI3}, the QPI map at $\bigl|\Delta_{\beta}\bigr|<\bigl|\Delta_{\alpha}\bigr|\approx\bigl|\Delta_{\gamma}\bigr|$ only reveals the intraband scattering (schematic) within the $\beta-$ FS pocket. (b2) All possible interband scatterings from $\alpha\rightarrow\gamma$-FS (${\bf q}_1$) and from $\gamma\rightarrow\gamma$-FS, (${\bf q}_{3,4}$), {\it but not from  $\beta\rightarrow\gamma$-FS (${\bf q}_2$)} become turned on at $\Omega^2_{\rm QPI}$.
(c1)-(c2) Due to inelastic scattering process in INS, the scattering between $\alpha$- or $\beta$- to $\gamma$-FSs is only allowed which changes sign of the SC gap, according to Eq.~\ref{eq:chi4}. ${\bf q}_2$ for $\beta\rightarrow\gamma$-FS appears at $\Omega^1_{\rm INS}=|\Delta_{\beta}|+|\Delta_{\gamma}|$ in (c1). (c2) ${\bf q}_1$ for $\alpha\rightarrow\gamma$-FS appears at $\Omega_{\rm INS}=|\Delta_{\alpha}|+|\Delta_{\gamma}|$.
As $|\Delta_{\alpha}|\approx|\Delta_{\gamma}|$, INS spectrum in (c2) will resemble QPI map in (b2) at zero magnetic field. Note that in the INS calculation, we have not included the Umklapp scattering which will symmetrize the INS spectra with respect to the Umklapp vector ${\bf Q}$. This is done to facilitate the direct comparison with QPI maps which are not calculated using Umklapp scattering to mimic the experimental procedure.}
\label{QPI_INS_pnictide}
\end{figure}

{\it FSs in pnictide:-}The low-energy Hamiltonian of Iron-pnictide system is dominated by five $d$-orbitals of the Fe atoms. We take the tight-binding model from Ref.~\onlinecite{maier} where the parameters are fit to the corresponding first-principles dispersion. The doping is evaluated within rigid-band shift approximation. At $x=-0.1$, the computed FS consists of two concentric hole pockets at $\Gamma$ points which are called $\alpha$- [inner pocket as depicted by red line in Fig.~1(a)] and $\beta$-pockets (blue line) and one electron pocket at M point (green line). All results in the this paper are presented in the unfolded BZ coming from 1 Fe unit cell.

{\it SC gap properties:-}From the shape of the FS pockets, there are at least four interpocket scattering channels exist in pnictide which span along various high-symmetry ${\bf q}$ directions as shown by arrows of different colors in Fig.~1(a). Among the four vectors, Mazin {\it et al.}\cite{Mazin} have shown theoretically that the nesting for ${\bf q}_{1}$ and ${\bf q}_2$ are the strongest which lead to a sign-changing $s^{\pm}$-pairing symmetry in this class of superconductors.  This phase symmetry of the SC gap is consistent with the spin-fluctuation mechanism of electron-pairing. The pairing and SC gap amplitudes have three essential properties which are relevant to our present study: (1) SC gap changes sign between the electron and hole pockets (black to white background colors in Fig.~1(a) reflect the $s^{\pm}$-pairing symmetry),  (2) SC gap magnitude on all FS pockets is nodeless and isotropic, and (3) evidence from ARPES\cite{HDing_twogaps}, STM\cite{NCYeh} and numerous bulk probes\cite{Johnston} indicate that,
\begin{equation}
|\Delta_{\alpha}|\approx|\Delta_{\gamma}|\approx 2|\Delta_{\beta}|,
\label{eq:gaps}
\end{equation}
at all dopings and for both electron and hole dopings.

{\it Sketch of QPI and INS maps:-} All the aforementioned FS and SC gap properties  lead to very different QPI and INS properties in pnictide, than the ones obtained in single band cuprates.\cite{Hanaguri_cuprate,Dastwogapcuprate}

(i)  Nodeless and isotropic nature of the SC gaps make all the QPI and INS maps to concentrate at only two energy scales, instead of a characteristic dispersion seen in nodal and anisotropic $d$-wave gap in cuprates. QPI maps will be prominent only at $\Omega^1_{\rm QPI}=\bigl|\Delta_{\beta}\bigr|$ and  $\Omega^2_{\rm QPI}=\bigl|\Delta_{\alpha}\bigr|\approx\bigl|\Delta_{\gamma}\bigr|$, obeying Eq.~\ref{eq:QPI3}.  While, the INS maps will show up only at $\Omega^1_{\rm INS}=\bigl|\Delta_{\beta}\bigr|+\bigl|\Delta_{\gamma}\bigr|$ and  $\Omega^2_{\rm INS}=\bigl|\Delta_{\alpha}\bigr|+\bigl|\Delta_{\gamma}\bigr|$, according to Eq.~\ref{eq:chi3}.

(ii) At $\Omega^1_{\rm QPI}=\bigl|\Delta_{\beta}\bigr|$, no interband elastic scattering is allowed as  $\bigl|\Delta_{\beta}\bigr|<\bigl|\Delta_{\alpha,\gamma}\bigr|$. Therefore, the QPI map, in Fig.~1(b1) only shows intra-$\beta$-FS scattering which concentrate near $q=0$. As SC gap does not change sign on each pocket, some finite magnetic field will be necessary to illuminate these small ${\bf q}$ vectors. Of course, in real material, finite broadening can introduce some quasiparticle states of $\alpha$, $\gamma$ bands near $\Omega_{\rm QPI}=\bigl|\Delta_{\beta}\bigr|$ to visualize weak intensity at ${\bf q}_2$ (at zero magnetic field).

(iii) At $\Omega^2_{\rm QPI}=\bigl|\Delta_{\alpha}\bigr|\approx\bigl|\Delta_{\gamma}\bigr|$, the interband scattering vector ${\bf q}_1$, intraband scattering vectors ${\bf q}_3$ and ${\bf q}_4$ appear in Fig.~1(b2).   ${\bf q}_3$ and ${\bf q}_4$ become illuminated at zero magnetic field while ${\bf q}_1$ will gain more intensity at finite magnetic field, in consistent with the data from Ref.~\onlinecite{Hanaguri_pnictide}.

(iv) At $\Omega^1_{\rm INS}=\bigl|\Delta_{\beta}\bigr|+\bigl|\Delta_{\gamma}\bigr|$, ${\bf q}_2$ vector can be observed through INS study in Fig.~1(c1). Note that, among the four maps shown in Fig.~1, this is the only place where ${\bf q_2}$ can be determined precisely.

(v) Similarly, at $\Omega^1_{\rm INS}=\bigl|\Delta_{\alpha}\bigr|+\bigl|\Delta_{\gamma}\bigr|$, ${\bf q}_1$ vector will be observed as sketched in Fig.~1(c2). At zero magnetic field, this INS spectra will match exactly with QPI pattern shown in Fig.~1(b2).

(vi) Doping dependence of the ${\bf q}$ vectors and their energy scales (not studied here) can also be used to gain confidence on the FS topology and the location of the sign reversal of the pairing symmetry. The INS energy scales, $\Omega^{1,2}_{\rm INS}$ obtain dome like behavior with doping in accord with the dome like behavior of the SC gaps as calculated in Ref.~\onlinecite{Dastworesonances}. By implication, the same doping dependence is expected in $\Omega^{1,2}_{\rm QPI}$. Even in a multi SC gaps pnictide system, such one-to-one correspondence is possible as all gaps show similar dome-like doping dependence.\cite{HDing_twogaps,NCYeh,Johnston} The area of the hole pockets increases with hole doping; simultaneously the same for the electron pocket decreases. This doping dependence is reversed for electron doping. Therefore, all ${\bf q}$ vectors will characteristically follow the doping dependence of their corresponding  connecting FSs.\cite{Dastworesonances} Interestingly, near $x=0.15$ of electron doped side, $\alpha$ pocket disappears.\cite{Dastworesonances,Terashima} Therefore, ${\bf q}_1$ should also disappear and hence $\Omega^2_{\rm QPI}$ (at zero magnetic field), $\Omega^2_{\rm INS}$. These results are yet to be confirmed experimentally.

\subsection{Computed QPI maps of iron-pnictide}

\begin{figure}
\rotatebox{0}{\scalebox{0.55}{\includegraphics{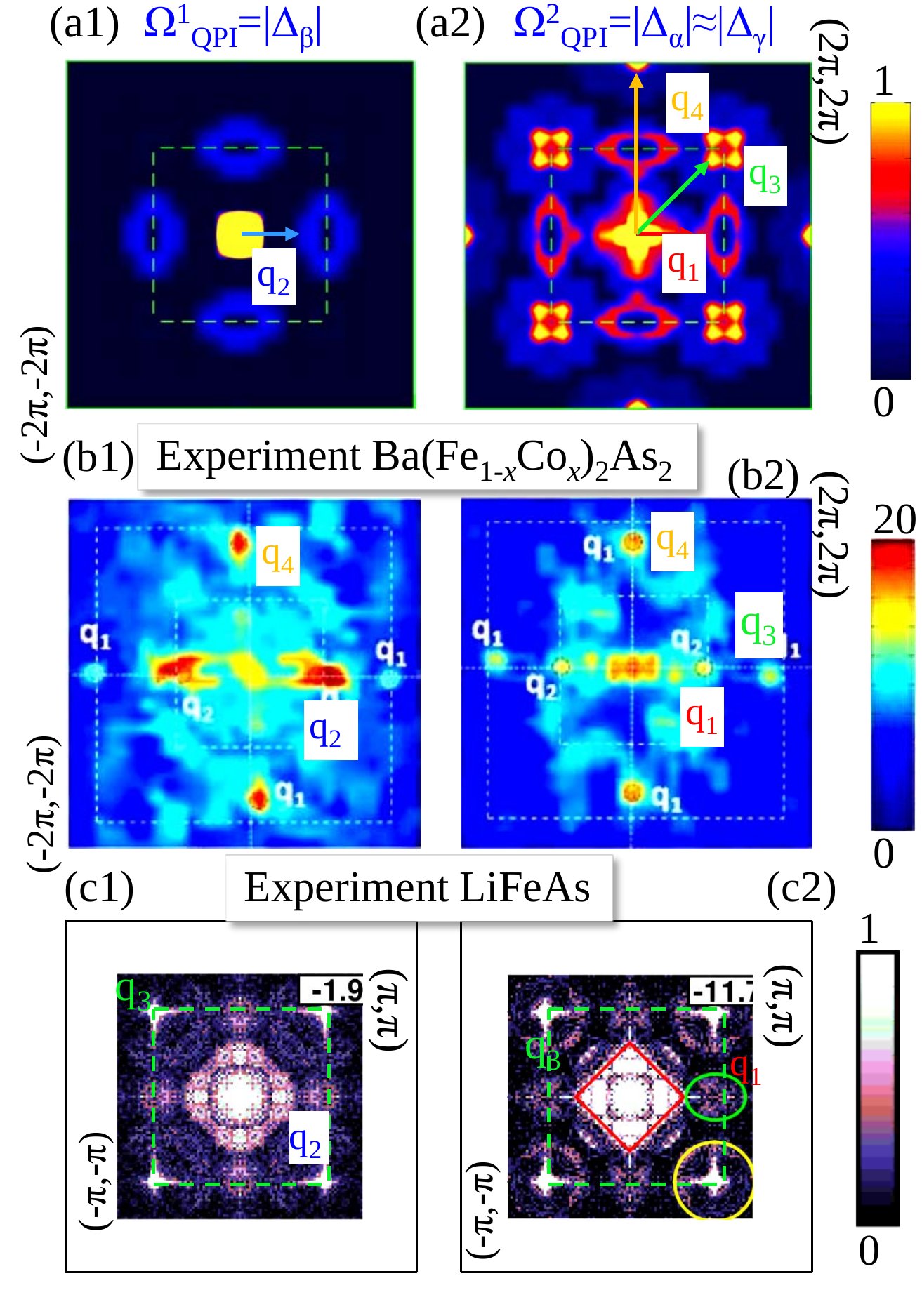}}}
\caption{(a1) Computed QPI map at  $\Omega^1_{\rm QPI}=|\Delta_{\beta}|$. As sketched in Fig.~1(b1), at this energy the ${\bf q}$ vectors come from intraband scattering within $\beta-$FS which concentrate near $q=0$. The weak intensity at ${\bf q}_2$ comes from finite broadening of the quasiparticle states at $\gamma$-pocket which allows some spectral weight of this band to appear around $\Omega^1_{\rm QPI}=|\Delta_{\beta}|$. (a2) Same as (a1) but at $\Omega^2_{\rm QPI}=|\Delta_{\alpha}|\approx|\Delta_{\gamma}|$. The results agree qualitatively with an earlier one,\cite{ZhangQPI} even after neglecting the orbital matrix element term in our case. However, we do not see the QPI pattern to be present apart from these two energy scales, unless a large lifetime broadening is used in the calculation to smear out the quasiparticle states. (b1)-(b2) Corresponding experimental QPI data at these two energy scales for electron doped pnictide at $x=0.06$ taken from Ref.~\onlinecite{NCYeh}. Experimental data show all four ${\bf q}$ vectors depicted in Fig.~1 at two energies, although their relative intensities are representative of our calculations in (a1) and (a2), respectively, due to low experimental resolution. As our phenomenological calculation do not capture the actual intensity of QPI maps, we normalize all QPI maps to their maximum. \blue{(c1)-(c2) Experimental data\cite{LiFeAsQPI} for a 111 compound LiFeAs at two bias energies $\Omega_{QPI}=$-1.9~meV in (c1), and  $\Omega_{QPI}=$-11.7~meV in (c2). Despise the differences in the FS topology for this 111 and our 122 system, we see that both results are in qualitative agreement. The experimental value of $\Delta$ for LiFeAs is 10~meV, which makes the result (c2) slightly above the SC region.\cite{LiFeAsQPI}}}
\label{QPI_pnictide}
\end{figure}

We compute the QPI maps for pnictide using Eq.~\ref{eq:QPI2} for the FSs given in Fig.~1(a). All results presented in Fig.~2 scales linearly with the SC gap magnitude as the relationship given in Eq.~\ref{eq:gaps} is maintained always. We will not include the scattering matrix-element $C$ [i.e., Eqs.~\ref{eq:QPI4} and \ref{eq:QPI5}] in this case to theoretically study the behavior of all ${\bf q}$ vectors as a function of energy. The computed results follow the similar behavior as sketched in Fig.~1, demonstrating that most of the evolution of the QPI maps can be understood from the simple energy and momentum conservation rules derived in Eqs.~\ref{eq:QPI3}- \ref{eq:QPI5}.

All the intraband scattering vectors lie so close to the strong elastic peak at ${\bf q}=0$ that it is often difficult to distinguish. We have taken broadening to be $\eta=1$~meV which is sufficient for the quasiparticle states at $\omega=|\Delta_{\gamma}|$ to extend up to $\omega=|\Delta_{\beta}|$, allowing some elastic scattering at ${\bf q}_2$ in Fig.~2(a1) (although it is prohibited in a clean limit). But our small broadening does not create visible intensity at all other interband vectors, although the experimental data at the corresponding energy shows some finite intensity at them, compare Fig.~2(a1) with corresponding experimental data in Fig.~2(b1).

At $\Omega^2_{\rm QPI}=|\Delta_{\alpha}|\approx|\Delta_{\gamma}|$, all interband scattering vectors between $\alpha$ to $\gamma$ FS appear on the QPI map as shown in Fig.~2(a2). As at $\Omega^1_{\rm QPI}$, ${\bf q}_2$ vector can only show up at $\Omega^2_{\rm QPI}$ due to residual broadening of the $\beta$ states upto $\gamma$ pocket. The separation between ${\bf q}_1$ and ${\bf q}_2$ can be studied more clearly in the overdoped region of hole doped side where the areas of the $\alpha$ and $\beta$ bands are distinguishably different.

The experimental data in Fig.~2(b2) also shows all the calculated ${\bf q}$ vectors. Subtle discrepancies in the relative intensities of each ${\bf q}$ vector is expected, because we have not included any matrix-element $M$, and the scattering coherence factor $C$ in this calculation. Furthermore, the magnitude of each ${\bf q}$ vectors do not match quantitatively with our calculation as the calculation is done at a different doping than the experimental data, although in both cases all three FS pockets are present. \blue{As both QPI and INS data are available as a function of momentum and energy for 111 family LiFeAs, we also compare our theory with these data. As shown in Fig.~2(c1) and 2(c2), the experimental data\cite{LiFeAsQPI} qualitatively agrees with the QPI vectors presented in the top panel. In performing such comparison, we have to pay attention to the difference in FS topology and SC gap symmetries between 122 and 111 families.}   

Once magnetic field is applied, the relative intensity of ${\bf q}_1$ with respect to that of ${\bf q}_{3,4}$ evolves  with the strength of the field, it has been observed in Fe(Se,Te) compounds, in accord with our calculations.\cite{Hanaguri_pnictide,knolleprl,knolleprb}

The proximity of ${\bf q}_4$ to the reciprocal vector $(2\pi,0)$ and its equivalent directions, which has been observed both in pnictide\cite{NCYeh} as well as in iron-chalcogenide\cite{Hanaguri_pnictide}, has been argued theoretically to arise from Bragg peak, instead of QPI scattering.\cite{Mazin_Singh} Comparing the evolution of the intensity of ${\bf q}_4$ at two energies $\Omega^1_{\rm QPI}$ in Fig.~2(a1) and  $\Omega^2_{\rm QPI}$ in Fig.~2(a2), we can deduce that it is not associated with Bragg peak as the latter does not have any energy dependence while ${\bf q}_4$ has.

\subsection{Computed INS maps of iron-pnictide}

\begin{figure}
\rotatebox{0}{\scalebox{0.47}{\includegraphics{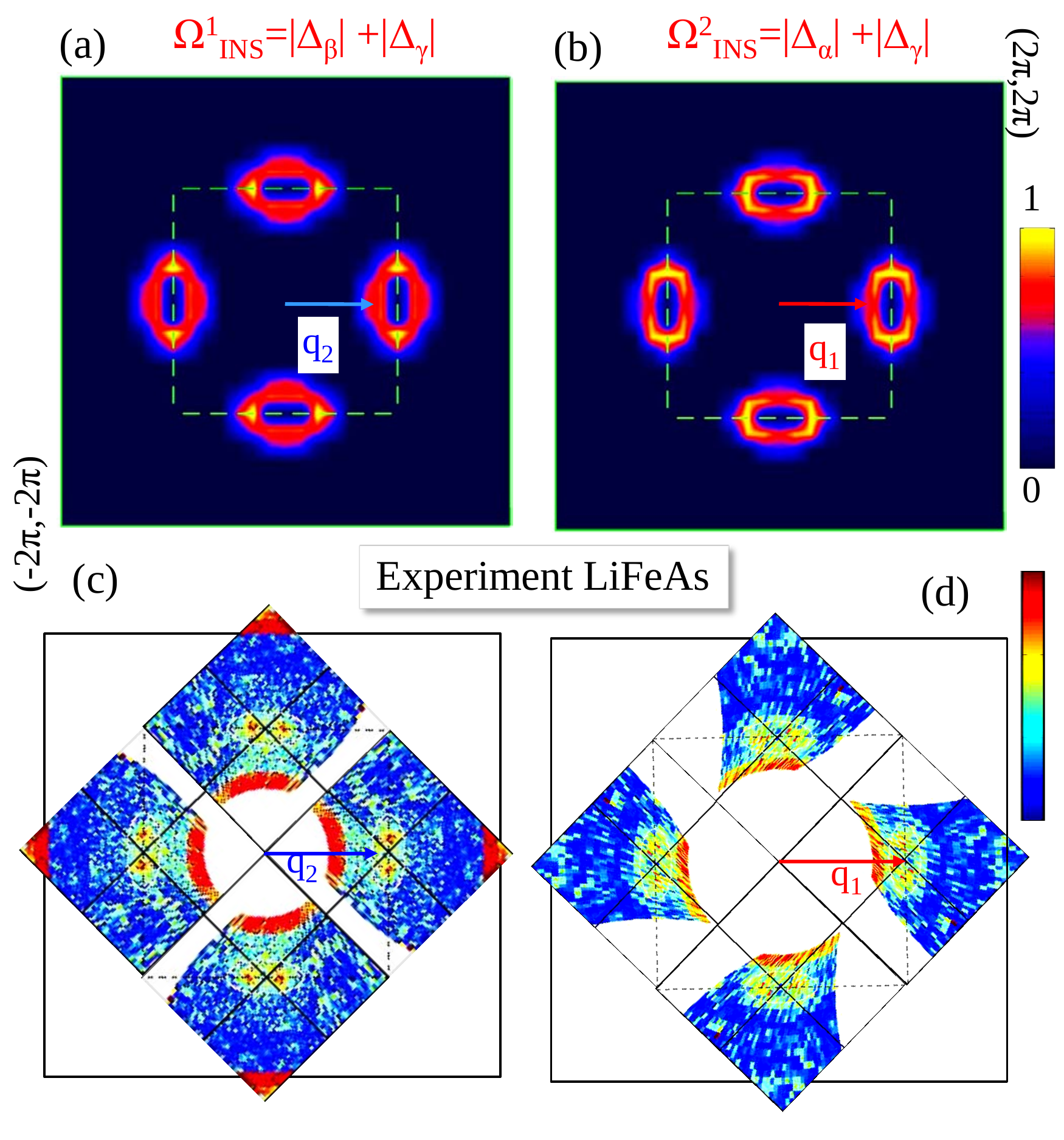}}}
\caption{(a) Computed INS map at  $\Omega^1_{\rm INS}=|\Delta_{\beta}|+|\Delta_{\gamma}|$ in 1 Fe unit cell notation. The intensity peak shifts away from $(\pi,0)$, implying that the corresponding resonance peak  is incommensurate due to the shape of the FSs. (b) Same as (a) but at $\Omega^1_{\rm INS}=|\Delta_{\alpha}|+|\Delta_{\gamma}|$. ${\bf q}_1$ peak is closer to the commensurate vector than ${\bf q}_2$ in (a) as $\alpha$-FS pocket is smaller than the $\beta$-one. The ${\bf q}$ values are strongly dependent on the FS areas, hence on doping. The results are consistent with the calculated spin-excitation dispersion plot presented in Ref.~\onlinecite{Dastworesonances} in 2 Fe unit cell. Both the incommensurate and the commensurate resonance peaks are observed in INS experiments.\cite{Christianson,Castellan_INS_INC} Note that the QPI maps shown in Figs.~2(a1) and 2(a2) will match exactly with (a) and (b), respectively, if a strong magnetic field is applied in the former case to eliminate the scattering of same sign of the SC gaps. \blue{(c)-(d) Experimental data for LiFeAs is presented at $\Omega_{INS}^1$=5~meV and $\Omega_{INS}^2$=10~meV from Ref.~\protect\onlinecite{LiFeAsINS}. The data is rotated by 45$^o$ as it is avaiable in 2 Fe unit cell notation, while the present calculation is performed in 1Fe unit cell notation.}}
\label{INS_pnictide}
\end{figure}

The INS spectra, a direct measure of $\chi^{\prime\prime}$ is calculated using Eq.~\ref{eq:chi2} and the results are shown in Figs.~3(a) and 3(b) at two energies where INS spectra is finite. The present phenomenological approach does not include the overlap matrix-element $M$ and the RPA correction. We have shown earlier in Ref.~\onlinecite{Dastworesonances} that they do not change the essential features of the INS for these systems, of course except intensity.

${\bf q}_2$ and ${\bf q}_1$ appear at $\Omega^1_{\rm INS}$ and $\Omega^2_{\rm INS}$, respectively as expected, because they involve sign change of the SC gaps between the initial and final states. ${\bf q}_2$ being smaller than ${\bf q}_1$, will lead to an incommensurate resonance while the latter is close to the commensurate one (both are doping dependence as described above). Both the commensurate\cite{Christianson} and the incommensurate\cite{Castellan_INS_INC} resonances have been detected by INS measurements, although the simultaneous presence of the two modes is yet to be detected in future measurements with better experimental resolution. \blue{The experimental data\cite{LiFeAsINS} for LiFeAs in Figs.~3(c) and 3(d) is consistent with the energy dependence of the ${\bf q}_2$ and ${\bf q}_1$ vectors.}

As mentioned earlier, the INS maps will correspond to the QPI maps if the latter is performed at zero magnetic field. \blue{For the experimental data of LiFeAs, such a compasion between Figs.~2(c1) and 2(c2) for QPI patterns and Figs.~3(c)-(d) for INS reveals good agreement.} 

\section{Layered iron-selenide}

\begin{figure}
\rotatebox{0}{\scalebox{0.65}{\includegraphics{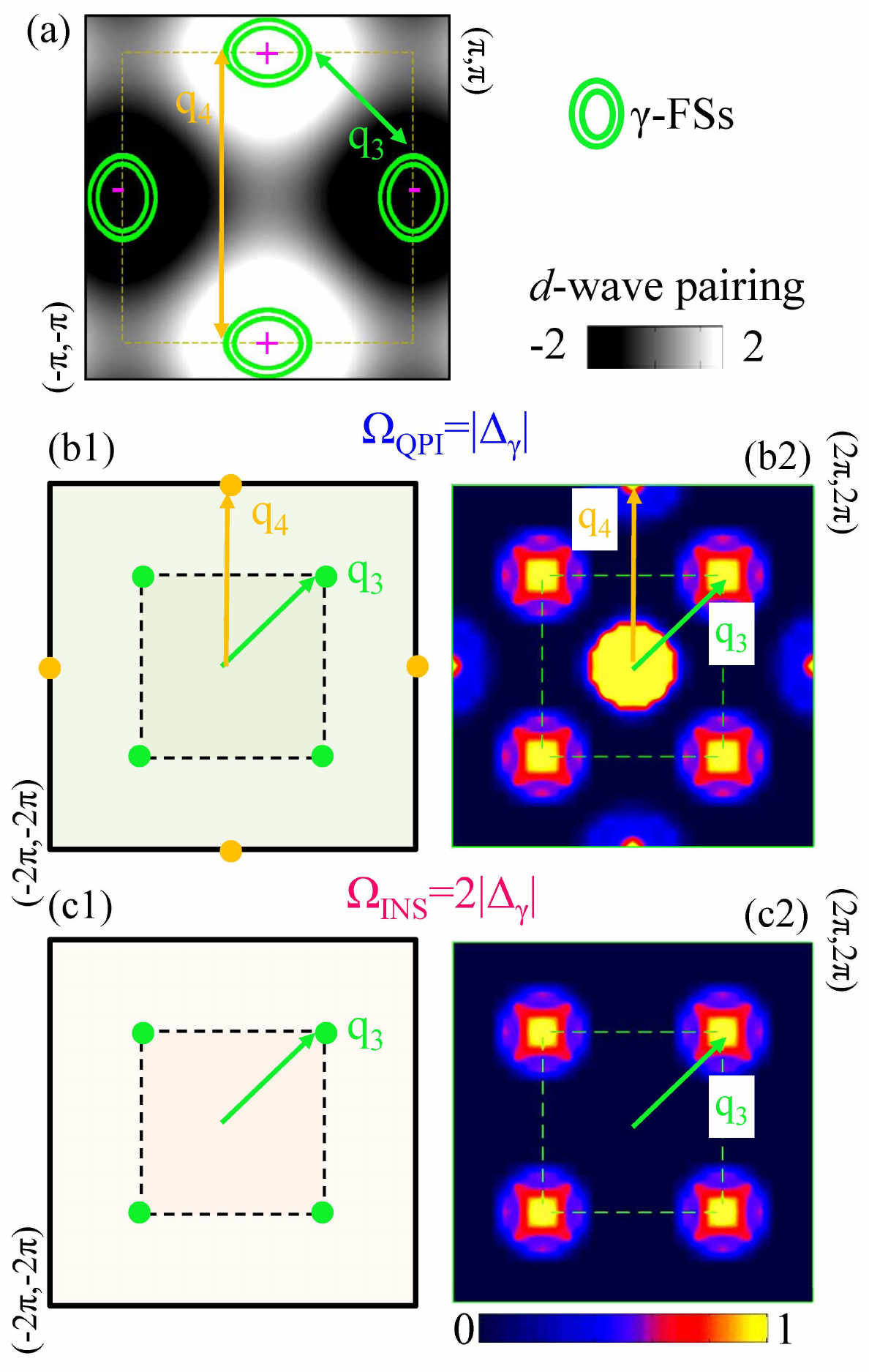}}}
\caption{(a) Computed FS for layered iron-selenide K$_x$Fe$_2$Se$_2$ system within two band tight-binding model at a representative in 1 Fe unit cell notation. The black to white background depicts the $d_{x^2-y^2}$-pairing symmetry which takes the form of $\cos{(k_xa)}-\cos{(k_ya)}$ in 1 Fe unit cell and $d_{xy}$-wave in 2 Fe unit cell (not shown).\cite{DasFeSe,Dasvacancy}
The arrows give two interband scattering channels that survive in this class of material, compared to iron-pnictide in Fig.~1(a). (b1) Sketch of the QPI map according to Eq.~\ref{eq:QPI3} (the associated scattering matrix-element which determines the sign of the SC state of the initial and final states is ignored here). (b2) Computed QPI map shows all ${\bf q}$-vectors predicted in (b1).  The small differences in the magnitude of ${\bf q}$-vectors from the two concentric electron pocket is not distinguishable due to finite broadening. (c1) In the INS spectra, ${\bf q}_4$ does not show up as it connects the quasiparticle state of same sign of SC gap. (c2) Corresponding computational result of INS spectra at $\Omega_{\rm INS}=2\Omega_{\rm QPI}$. The resonance spectra is commensurate at $(\pi,\pi)$, while the inclusion of orbital matrix-element $M$ shifts it to a slightly incommensurate one at a critical value of the interaction $U$.\cite{DasFeSe} Our prediction agrees with recent INS measurements in this class of materials.\cite{Inosov_INS_FeSe} }
\label{QPI_INS_FeSe}
\end{figure}

{\it FS properties:-}We turn next to the layered $A_y$Fe$_{2-x}$Se$_2$ based systems. These materials host only electron pockets at M point with no hole pocket.\cite{HDing_FeSe} To model such electronic structure, we employ two band tight-binding calculation of the $t_{2g}$ orbitals of Fe $d$ bands and the parameters are obtained by fitting to the material specific LDA dispersion.\cite{DasFeSe} The resulting FSs are shown by green lines in Fig.~4(a), which match well with ARPES FS.\cite{HDing_FeSe}

{\it SC gap properties:-}The absence of hole pocket results in strong nesting between the electron pockets along ${\bf q}_3$. We have shown earlier that such nesting lead to nodeless and isotropic $d_{x^2-y^2}-$ pairing symmetry [in 2 Fe unit cell the pairing symmetry becomes $d_{xy}$\cite{Dasvacancy}] and a spin resonance near the commensurate vector ${\bf q}_3$.\cite{DasFeSe} The spin-resonance has recently been found experimentally in this class of material by Park {\it et al.}\cite{Inosov_INS_FeSe}.

{\it QPI and INS spectra:-}Due to one SC gap and one FS (concentric electron pockets), both the QPI and INS maps appear only at one energy scale. The QPI map in this material as shown in Fig.~4(b1) resembles the QPI map for pnictide in Fig.~2(b2), with the exception that the electron-hole scattering ${\bf q}_2$ is absent here. On the other hand, the INS spectra [Fig.~4(c)] is rotated by 45$^o$ in comparison with the same plot of pnictide in Fig.~3(b) as the `hot-spot' for sign-reversal SC gap is now aligned along ${\bf q}_3$. It is interesting to note that although ${\bf q_3}$ scattering is also present in pnictide as seen in QPI maps, but the leading nesting shifts along ${\bf q}_1$, giving a very different pairing symmetry. This observation leads us to conclude that subtle differences between the nesting and the scattering which can carry rich physical insight can be untangled more   clearly by comparing the QPI with INS spectra.

\subsection{Magnetic field dependent QPI maps}

\begin{figure}
\rotatebox{0}{\scalebox{0.35}{\includegraphics{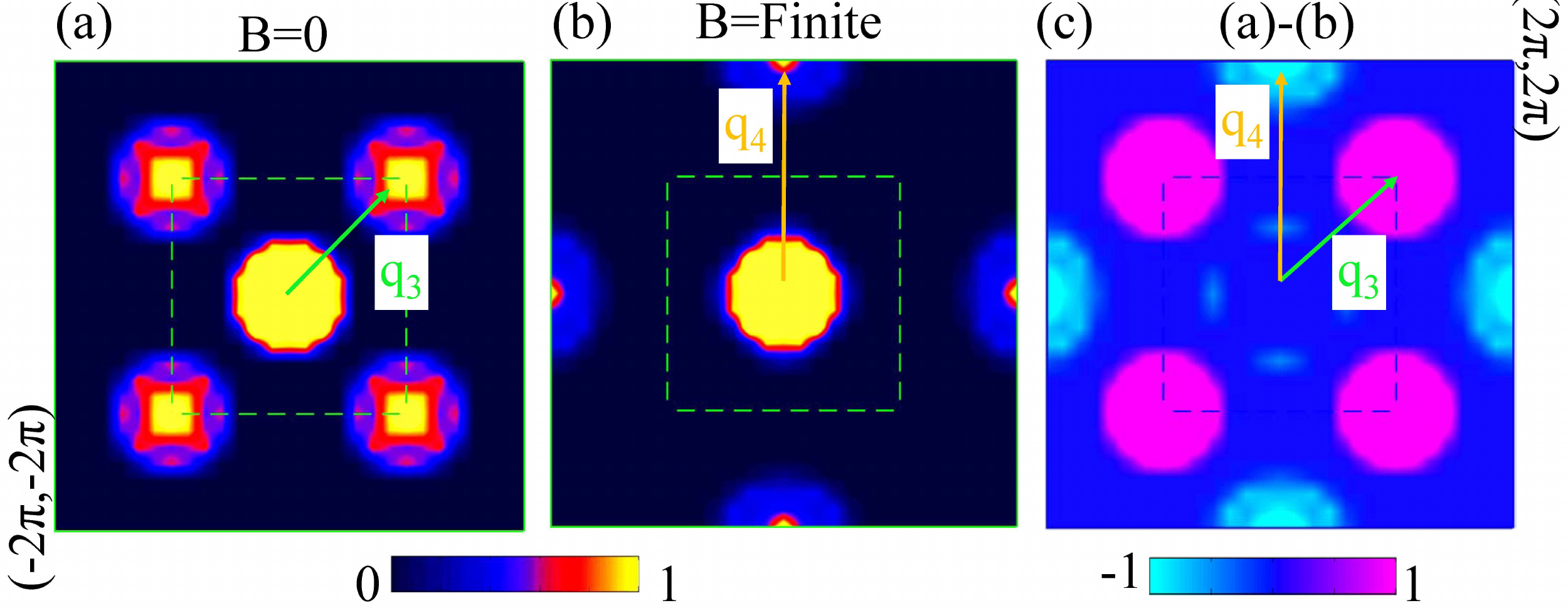}}}
\caption{(a) Computed result of QPI pattern in K$_x$Fe$_2$Se$_2$ using Eqs.~\ref{eq:QPI3} and \ref{eq:QPI4} which mimics the zero magnetic field ($B$) condition at which those QPI vectors become illuminated which correspond to the scattering of states having opposite sign of the SC gaps.\cite{Hanaguri_cuprate} (b) Same as (a) but using Eqs.~\ref{eq:QPI3} and \ref{eq:QPI5}. This means, scatterings of same sign of the quasiparticle states are only included which mimic a finite applied magnetic field condition. In principle, a magnetic impurity has a scalar component which will allow some scattering for ${\bf q}_3$ vector as well but relatively weak in intensity. (c) Differences between (a) and (b). The positions of maxima and minima corresponds to ${\bf q}_3$ and ${\bf q}_2$, respectively. an unambiguous method to find out which scattering channel involve sign change of the SC gap, a test which will confirm the presence of $d$-wave gap in these materials.}
\label{QPI_B_FeSe}
\end{figure}

As mentioned before, another way to distinguish the scattering and nesting vector in QPI map is to study its  magnetic field dependence of it at fixed energy. At zero magnetic field, the scattering matrix element $C$ only allows scattering of quasiparticle states of opposite phase of the SC gap, see Eq.~\ref{eq:QPI4}. Therefore, the resulting QPI map at $\Omega_{\rm QPI}$ will resembles the INS spectra at $\Omega_{\rm INS}=2\Omega_{\rm QPI}$, although the ${\bf q}=0$ elastic peak will be difficult to remove in the former case.

At finite magnetic field, the other scattering channel, ${\bf q}_4$, at which the scattering from states of same sign of the SC gap becomes turned on as shown in Fig.~5(b). In practice, a magnetic impurity is always associated with a scalar component (not considered in the present calculation for simplicity), therefore, ${\bf q}_3$ should also be present. The relative intensity can be monitored by tuning the strength of the magnetic field.\cite{Hanaguri_cuprate,Hanaguri_pnictide} On the otherhand, the difference between QPI maps in Fig.~5(a) and 5(b), as shown in Fig.~5(c), can also be used to identify the sign-changing `hot-spot' vector. The positions of maxima and minima corresponds to ${\bf q}_3$ and ${\bf q}_4$, respectively. The experimental detection of the maxima and minima can be tested to confirm the presence of $d$-wave gap in these materials.

\section{conclusions} We have developed a phenomenological approach that allows one to connect the INS and QPI maps. We point out that both the INS and QPI maps arise due to the inelastic and elastic scattering of the Bogoliubov quasiparticles, respectively, and bear a one-to-one correspondence  as a function of scattering vector and energy. In fact, inelastic and elastic scattering complements these two spectroscopies to quantitatively and unambiguously identify the nature of pairing symmetry in unconventional superconductors. By applying our model in iron-pnictide and layered iron selenide compounds, we show that the QPI maps at zero magnetic field corresponds exactly to the INS spectra at their representative energy (the $q=0$ elastic scattering and other spurious effects in QPI maps can be ignored because of ``contamination from Bragg peaks''). We also show that upon applying magnetic field , the QPI scattering at same sign of the SC gaps can be illuminated in layered iron-selenides and produce detectable changes in QPI. We point that evolution of the QPI maps can be implemented  experimentally to test the  possible nodeless $d$-wave pairing in this class of materials.

\begin{acknowledgments}
We are grateful to Y. K. Bang, T. Hanaguri, H. Takagi, J.-X. Zhu, H. H. Wen, H. Ding, N. C. Yeh for useful discussions.
This  work was supported by This work was supported by the U.S. DOE at Los Alamos
National Laboratory under contract No. DE-AC52-06NA25396 and the Office of Science (BES) and benefited from the NERSC computing allocations.
\end{acknowledgments}


\begin{thebibliography}{99}
\bibitem{strongcoupling} In a strong coupling scenario, the mechanism for pairing may be different. However, our QPI and INS results will be similar as long as the pairing symmetry and the Fermi surface topology remain same. See, for example, W. Q. Chen, Phys. Rev. Lett. {\bf 102}, 047006 (2009); and C. Fang {\it et al.}, Phys. Rev X {\bf 1}, 011009 (2011), for strong coupling theory of the pairing symmetry.
%
\bibitem{scalapino}D. J. Scalapino, E. Loh, Jr., and J. E. Hirsch, Phys. Rev. B {\bf 35}, 6694 (1987).
%
\bibitem{Pines}P. Monthoux, A. V. Balatsky, and D. Pines, Phys. Rev. Lett. {\bf 67}, 3448 (1991).
%
\bibitem{Mazin} I. I. Mazin, D. J. Singh, M. D. Johannes, M. H. Du, Phys. Rev. Lett. {\bf 101}, 057003 (2008).
%
\bibitem{DHLeeFeSe}Fa Wang, F. Yang, M. Gao, Z.-Yi Lu, T. Xiang, D.-H. Lee, Europhy. Lett. {\bf 93}, 57003 (2011) .
%
\bibitem{MaierFeSe}T. A. Maier, S. Graser, P. J. Hirschfeld, D. J. Scalapino, Phys. Rev. B {\bf 83}, 100515(R) (2011) .
%
\bibitem{DasFeSe} T. Das, and A. V. Balatsky, Phys. Rev. B {\bf 84}, 014521 (2011).
%
\bibitem{Dasvacancy} T. Das, and A. V. Balatsky, Phys. Rev. B {\bf 84}, 115117 (2011).
%
\bibitem{chubukov}Ar.  Abanov, and A. V. Chubukov, Phys. Rev. Lett. {\bf 83}, 165 (1999).
%
\bibitem{norman} M. Eschrig, and M.. R.  Norman,
%
Phys. Rev. B {\bf 67}, 1445031 (2003).
%
\bibitem{eremin}I. Eremin, D. K. Morr, A. V. Chubukov, K. H. Bennemann, M. R. Norman, Phys. Rev. Lett. {\bf 94}, 147001 (2005).
%
\bibitem{Dastworesonances} T. Das, and A. V. Balatsky, Phys. Rev. Lett. {\bf 106}, 157004 (2011).
%
\bibitem{so5} E. Delmer, and S.-C. Zhang, Phys. Rev. Lett. {\bf 75}, 4126 (1995).
%
\bibitem{PLee_slaveboson}J. Brinckmann, and P. A. Lee, Phys. Rev. Lett. {\bf 82}, 2915 (1999).
%
\bibitem{wilsonPRB}F. Kr\"uger, S. D. Wilson, L. Shan, S. Li, Y. Huang, H.- H. Wen, S.-C. Zhang, P. Dai, J. Zaanen, Phys. Rev. B {\bf 76}, 094506 (2007).
%
\bibitem{Hanaguri_cuprate}T. Hanaguri, Y. Kohsaka, M. Ono, M. Maltseva, P. Coleman, I. Yamada, M. Azuma, M. Takano, K. Ohishi, H. Takagi, Science {\bf 323}, 923 (2009).
%
\bibitem{Hanaguri_pnictide} T. Hanaguri, S. Niitaka, K. Kuroki, H. Takagi, Science {\bf 328}, 474 (2010).
%
\bibitem{knolleprl}J. Knolle, I. Eremin, A. Akbari, and R. Moessner, Phys. Rev. Lett. {\bf 104}, 257001 (2010).
%
\bibitem{knolleprb} A. Akbari, J. Knolle, I. Eremin, and R. Moessner, Phys. Rev. B {\bf 82}, 224506 (2010).
%
\bibitem{Schrieffer_book} J. R. Schrieffer, Theory of Superconductivity (Perseus, Reading, MA, 1999).
%
\bibitem{electron_pocket}T. Das, R. S. Markiewicz, A. Bansil, Phys. Rev. B {\bf 85}, 064510 (2012).
%
\bibitem{HDing_twogaps}K. Nakayama, T. Sato, P. Richard, Y.-M. Xu, T. Kawahara, K. Umezawa, T. Qian, M. Neupane, G. F. Chen, H. Ding, and T. Takahashi, Phys. Rev. B {\bf 83}, 020501(R) (2011).
%
%
\bibitem{Johnston}D. C. Johnston, Adv. in Phys. {\bf 59}, 803 (2010).
%
\bibitem{Hoffman} J. Hoffman, Science {\bf 328}, 441 (2010).
%
\bibitem{NCYeh}M. L. Teague, G. K. Drayna, G. P. Lockhart, P. Cheng, B. Shen, H.-H. Wen, and N.-C. Yeh, Phys. Rev. Lett. {\bf 106}, 087004 (2011).
%
\bibitem{FeSe} J. Guo, S. Jin, G. Wang, S. Wang, K. Zhu, T. Zhou, M. He, and X. Chen, Phys. Rev. B {\bf 82}, 180520(R) (2010).
%
\bibitem{FeSe1} A. Krzton-Maziopa, Z. Shermadini, E. Pomjakushina, V. Pomjakushin, M. Bendele, A. Amato, R. Khasanov, H. Luetkens, and K. Conder,  J. Phys.: Condens. Matter {\bf 23}, 052203 (2011).
%
\bibitem{FeSe2} M. Fang, H. Wang, C. Dong, Z. Li, C. Feng, J. Chen, and H.Q. Yuan, Europhys. Letts. {\bf 94}, 27009 (2011).
%
%
\bibitem{Zhang}Y. Zhang, L. X. Yang, M. Xu, Z. R. Ye, F. Chen, C. He, J. Jiang, B. P. Xie, J. J. Ying, X. F. Wang, X. H. Chen, J. P. Hu, and D. L. Feng, Nature Materials {\bf 10}, 273 (2011).
%
\bibitem{HDing} X.-P. Wang, T. Qian, P. Richard, P. Zhang, J. Dong, H.-D. Wang, C.-H. Dong, M.-H. Fang, and H. Ding,
Europhysics Letts. {\bf 93}, 57001 (2011).
\bibitem{Mou} D. Mou,S. Liu, X. Jia, J. He, Y. Peng, L. Zhao, L. Yu, G. Liu, S. He, X. Dong, J. Zhang, H. Wang, C. Dong, M. Fang, X. Wang, Q. Peng, Z. Wang, S. Zhang, F. Yang, Z. Xu, C. Chen and X. J. Zhou,
Phys. Rev. Lett. {\bf 106}, 107001 (2011).
%
\bibitem{Cv}B. Zeng, B. Shen, G. Chen, J. He, D. Wang, C. Li, and H.-Hu Wen, Phys. Rev. B {\bf 83}, 144511 (2011).
%
\bibitem{Inosov_INS_FeSe}J. T. Park, G. Friemel, Yuan Li, J.-H. Kim, V. Tsurkan, J. Deisenhofer, H.-A. Krug von Nidda, A. Loidl, A. Ivanov, B. Keimer, and D. S. Inosov, Phys. Rev. Lett. {\bf 107}, 177005 (2011).
%
\bibitem{SashaQPI}J.-X. Zhu, K. McElroy, J. Lee, T. P. Devereaux, Q. Si, J. C. Davis, and A. V. Balatsky, Phys. Rev. Lett. {\bf 97}, 177001 (2006).
%
\bibitem{BobQPI}R. S. Markiewicz, Phys. Rev. B {\bf 69}, 214517 (2004).
%
\bibitem{Dastwogapcuprate}T. Das, R. S. Markiewicz, and A. Bansil, Phys. Rev. B {\bf 77}, 134516 (2008).
%
\bibitem{ZhangQPI}Y.-Y. Zhang, C. Fang, X. Zhou, K. Seo, W.-F. Tsai, B. A. Bernevig, J. Hu, Phys. Rev. B {\bf 80}, 094528 (2009).
%
\bibitem{maier}T. A. Maier, S. Graser, D. J. Scalapino, and P. Hirschfeld, Phys. Rev. B {\bf 79}, 134520 (2009).
%
\bibitem{Terashima} K. Terashima, Y. Sekibab, J. H. Bowenc, K. Nakayamab, T. Kawaharab, T. Satob, P. Richarde, Y.-M. Xuf,
L. J. Lig, G. H. Caog, Z.-A. Xug, H. Dingc, and T. Takahashi, Proc. Natl. Acad. Sci. U.S.A. {\bf 106}, 7330 (2009).
%
%
\bibitem{LiFeAsQPI}T. HÃ¤nke, S. Sykora, R. Schlegel, D. Baumann, L. Harnagea, S. Wurmehl, M. Daghofer, B. BÃ¼chner, J. van den Brink, C. Hess, arXiv:1106.4217.
%
\bibitem{Mazin_Singh}I. I. Mazin, and D. J. Singh, e-print arXiv:1007.0047.
%
\bibitem{Christianson}A. D. Christianson, E. A. Goremychkin, R. Osborn, S. Rosenkranz, M. D. Lumsden, C. D. Malliakas, I. S. Todorov, H. Claus, D. Y. Chung, M. G. Kanatzidis, R. I. Bewley, and T. Guid, Nature {\bf 456}, 930 (2008).
%
\bibitem{Castellan_INS_INC}J.-P. Castellan, S. Rosenkranz, E. A. Goremychkin, D. Y. Chung, I. S. Todorov, M. G. Kanatzidis, I. Eremin, J. Knolle, A. V. Chubukov, S. Maiti, M. R. Norman, F. Weber, H. Claus, T. Guidi, R. I. Bewley, and R. Osborn, Phys. Rev. Lett. {\bf 107}, 177003 (2011).
%
\bibitem{LiFeAsINS}N. Qureshi, P. Steffens, Y. Drees, A. C. Komarek, D. Lamago, Y. Sidis, L. Harnagea, H.-J. Grafe, S. Wurmehl, 5 B. BÃ¼chner, M. Braden, Phys. Rev. Lett. {\bf 108}, 117001 (2012).
%
%
\bibitem{HDing_FeSe} X.-P. Wang, T. Qian, P. Richard, P. Zhang, J. Dong, H.-D. Wang, C.-H. Dong, M.-H. Fang, and H. Ding,
Europhysics Letts. {\bf 93}, 57001 (2011).
%
\end{thebibliography}
\end{document}